\documentclass[onecolumn,aps,pra,superscriptaddress,groupedaddress,notitlepage,nofootinbib]{revtex4-1}

\usepackage{amssymb,amsmath,color,graphicx}
\usepackage{hyperref}
\usepackage{enumerate}
\usepackage{subfigure}
\usepackage{dsfont}
\usepackage{bbold}

\newcommand{\bra}[1]{\langle{#1}|}
\newcommand{\ket}[1]{|{#1}\rangle}

\begin{document}

\date{July 17, 2017}

\title{Optimal secure quantum teleportation of coherent states of light}

\author{Pietro Liuzzo-Scorpo}
\author{Gerardo Adesso}
\thanks{%Further author information: (Send correspondence to G.A.)\\G.A.:
E-mail: gerardo.adesso@nottingham.ac.uk}
\affiliation{
Centre for the Mathematics and Theoretical Physics of Quantum Non-Equilibrium Systems, \\ Quantum Correlations Group, School of Mathematical Sciences, University of Nottingham, \\
University Park, Nottingham NG7 2RD, United Kingdom}

%\authorinfo{$^{\dagger}$Further author information: (Send correspondence to G.A.)\\G.A.: E-mail: gerardo.adesso@nottingham.ac.uk}

\begin{abstract}

We investigate quantum teleportation of ensembles of coherent states of light with a Gaussian distributed
displacement in phase space. Recently, the following general question has been addressed in [P.~Liuzzo-Scorpo {\it et al.}, 
Phys. Rev. Lett. {\bf 119}, 120503 (2017)]: Given a limited amount of
entanglement and mean energy available as resources, what is the maximal fidelity that can be achieved on average in the
teleportation of such an alphabet of states? Here, we consider a variation of this question, where Einstein--Podolsky--Rosen steering is used as a resource rather than plain entanglement. We provide a solution by means of an optimisation within the space of Gaussian quantum channels, which allows for an intuitive visualisation of the problem. We
first show that not all channels are accessible with a finite degree of steering, and then prove that practical
schemes relying on asymmetric two-mode Gaussian states enable one to reach the
maximal fidelity at the border with the inaccessible region. Our results provide a rigorous quantitative assessment of steering as a resource for secure quantum teleportation beyond the so-called no-cloning threshold. The schemes we propose can be readily implemented experimentally by a conventional Braunstein--Kimble continuous variable teleportation
protocol involving homodyne detections and corrective displacements with an optimally tuned gain. These
protocols can be integrated as elementary building blocks in quantum networks, for reliable storage and
transmission of quantum optical states.
\end{abstract}

\maketitle

\section{Introduction}

The quantum internet is dawning \cite{qinternet}. A confluence of technologies, including quantum optics, nano-manufacturing, fibre-based and satellite telecommunications, are accelerating the realisation of a large-scale quantum communication network. The operation of such a network is relying on key primitives, such as storage, retrieval, and teleportation of quantum states of light. In particular, {\it teleportation} \cite{Bennett1993} enables the transfer of unknown quantum states from a sender (Alice) to a receiver (Bob) without physical transport of their carriers, thanks to shared quantum resources such as entanglement distributed over a quantum channel between Alice and Bob. It is of fundamental and practical importance to optimise the performance of teleportation and related communication protocols, while keeping the costs involved in the preparation of the shared resources to a minimum \cite{Liuzzo2017}.

There are several ways to benchmark the success of a teleportation protocol. Suppose an input state $\ket{\psi_z}^{\rm in}$, unknown to Alice, is drawn from a preassigned set  according to a prior probability distribution $p(z)$ known to Alice and Bob. After a run of the protocol,  the corresponding output state obtained by Bob can be denoted as $\rho_z^{\rm out}$ and its {\it fidelity} with the unknown input, as measurable by an independent verifier (Victor), is given by ${\cal F}(z) = {}^{\rm in}\bra{\psi_z}\rho^{\rm out}\ket{\psi_z}^{\rm in}$. A suitable figure of merit quantifying the performance of teleportation in this setting is given by the average fidelity over the input ensemble, defined as
\begin{equation}\label{eq:avf}
\bar{\cal F} = \int_z dz \ p(z)\  {\cal F}(z)\,.
\end{equation}

In particular, teleportation of an input alphabet of quantum states is {\it certified quantum} when the average fidelity $\bar{\cal F}$ exceeds a threshold $\bar{\cal F}^{(c)}$ corresponding to the best classical scheme, where Alice attempts to directly measure each input state and then requests Bob to prepare an output state based on her communicated outcome \cite{Braunstein2000}. However, surpassing this threshold does not exclude that a malicious eavesdropper (Eve) trying to interfere with the communication might receive a copy of the transferred state which scores a better fidelity than Bob's output with Alice's input. Teleportation is then {\it certified secure} when the average fidelity $\bar{\cal F}$  exceeds a larger threshold  $\bar{\cal F}^{(s)}$ obtained by averaging the fidelity of the best   $1 \to 2$ cloning protocol over the input ensemble \cite{GG}.

Achieving security in quantum teleportation is an essential requirement for a reliable quantum internet. In practice, though, reaching $\bar{\cal F} > \bar{\cal F}^{(s)}$ may require significantly more resources than just fulfilling $\bar {\cal F} > \bar{\cal F}^{(c)}$, depending on the specifics of the input set. In fact, comparatively very little is known about the exact nature of the resources needed for certified secure teleportation, as opposed to the case of `only' certified quantum teleportation, which is well known to rely on entanglement \cite{NP2015}. A recent work \cite{He2015} has made some progress towards this question, considering in particular the secure continuous variable teleportation of an input alphabet of coherent states of light \cite{Glauber1963} $\ket{\alpha}^{\rm in}$  with uniform distribution in phase space. In such a case, Einstein-Podolsky-Rosen (EPR) steering, a form of asymmetric nonlocality stronger than entanglement \cite{Wiseman2007}, has been identified as the necessary resource to attain the secure teleportation regime, which amounts to reaching ${\bar{\cal F}} > 2/3$ \cite{GG}.

Here we will address the question in more general terms, and establish the optimal average fidelity $\bar{\cal F}$  for teleporting an alphabet of coherent states with non-uniform (Gaussian) phase space prior distribution, given a finite amount of EPR steering available as a resource between Alice and Bob. This provides a feasible prescription of immediate practical relevance for the realisation of secure continuous variable teleportation of ensembles of states of light in quantum optics.

In continuous variable teleportation \cite{Braunstein2005}, Alice can in principle teleport any input state to Bob with unit fidelity if they share an ideal EPR state \cite{EPR35,Vaidman1994}. However, such a state has infinite mean energy and is therefore unphysical, yet it can be approximated arbitrarily well by families of resource states with finite mean energy. In particular, Gaussian states of light \cite{Weedbrook2012,Adesso2014,Serafini2017}, such as two-mode squeezed thermal states, are valuable resources for continuous variable teleportation with limited resources, according to the protocol proposed by Braunstein and Kimble (BK) \cite{Braunstein1998}.

In the following, we will briefly review the basics of Gaussian quantum information theory (Sec.~\ref{sec:cv}), including states, channels, separability and steerability, as well as the BK teleportation protocol. We will then characterise the single-mode Gaussian channels which can be simulated by teleportation exploiting a limited steering resource (Sec.~\ref{sec:sim}) and derive the optimal average fidelity for secure teleportation of coherent states of light as a function of the available steering and the input distribution variance (Sec.~\ref{sec:opt}).

\section{Basics of Gaussian quantum information theory}\label{sec:cv}

\subsection{Gaussian states}

Given a $n$-mode continuous variable system, described by a vector of canonical operators $\hat{R}=(\hat{q}_1, \hat{p}_1,\dots,\hat{q}_n,\hat{p}_n )^\top$, Gaussian states are entirely specified by their first and second moments \cite{Weedbrook2012,Adesso2014,Serafini2017}, given respectively by the displacement vector   $d=\langle\hat{R}\rangle$ and the covariance matrix $V$ with entries $V_{jk}=\langle\{\hat{R}_j-d_j,\hat{R}_k-d_k\}_+\rangle$. For $n$-mode Gaussian states with vanishing displacement vector, the mean energy (precisely, the mean photon number $\bar n$) per mode can be computed from the covariance matrix via the formula
\begin{equation}\label{eq:energy}
\bar{n} = \frac{1}{4n} (\text{tr}\, V - 2n)\,.
\end{equation}

Specialising to two modes $A$ and $B$ ($n=2$), a bipartite Gaussian state with covariance matrix $V_{AB}$ is:
\begin{itemize}
\item physical if and only if \cite{Simon1994}
\begin{equation}\label{eq:bonafide}
V_{AB} + i (\omega_A \oplus  \omega_B) \geq 0\,,
\end{equation}
where $\omega = \left(
\begin{array}{cc}
 0 & 1 \\
 -1 & 0 \\
\end{array}
\right)$ is the symplectic matrix encoding the canonical commutation relations for each mode;
\item
separable if and only if \cite{Simon00,WerWolf00}
\begin{equation}\label{eq:separable}
V_{AB} + i (- \omega_A \oplus  \omega_B) \geq 0\,;
\end{equation}
\item
not steerable by Gaussian measurements from $B$ to $A$ if and only if \cite{Wiseman2007}
\begin{equation}\label{eq:unsteerableba}
V_{AB} + i (\omega_A  \oplus \mathbb{0}_B) \geq 0\,;
\end{equation}
\item
not steerable by Gaussian measurements from $A$ to $B$ if and only if \cite{Wiseman2007}
\begin{equation}\label{eq:unsteerableab}
V_{AB} + i (\mathbb{0}_A  \oplus \omega_B) \geq 0\,.
\end{equation}
\end{itemize}
Violation of (\ref{eq:separable}) amounts to entanglement between the modes $A$ and $B$, while violation of (\ref{eq:unsteerableba}) [resp.~(\ref{eq:unsteerableab})] reveals $B \to  A$ (resp.~$A  \to B$) steering, a manifestation of the EPR paradox \cite{reid} which enables entanglement verification even if Bob's (resp.~Alice's) devices are untrusted or generally uncharacterised \cite{Wiseman2007}.

In block form, any two-mode covariance matrix can be written as
\begin{equation}\label{eq:cm}
V_{AB}=\left(\begin{array}{c|c}
A & C \\ \hline
C^\top  & B
\end{array}\right)\,,
\end{equation}
where $A$ and $B$ are the reduced covariance matrices of Alice's and Bob's modes, respectively, and the off-diagonal block $C$ contains correlations between the modes. The bona fide condition (\ref{eq:bonafide}) can be reformulated as a condition on the symplectic spectrum of the covariance matrix,
\begin{equation}
\label{eq:bonafidenu}
\nu^+_{AB} \geq \nu^-_{AB} \geq 1\,,
\end{equation}
where the symplectic eigenvalues $\nu^{\pm}_{AB}$ are defined by \cite{Adesso2004}
\begin{equation}
\label{eq:nupm}\nu^{\pm}_{AB} =
\sqrt{\frac12 \left(\Delta_{AB} \pm \sqrt{\Delta_{AB}^2 - 4 \det V_{AB}}\right)}\,,
\end{equation}
with $\Delta_{AB} = \det A + \det B + 2 \det C$.

With respect to the block form of Eq.~(\ref{eq:cm}), the steerability of a Gaussian state with covariance matrix $V_{AB}$ by Gaussian measurements in either direction can be quantified by \cite{Kogias2014}
\begin{eqnarray}
{\cal S}_{B \to A}(V_{AB})&=&\max\left\{0,\frac12\log\left(\frac{\det B}{\det V_{AB}}\right)\right\}\,, \label{eq:sba}\\
{\cal S}_{A \to B}(V_{AB})&=&\max\left\{0,\frac12\log\left(\frac{\det A}{\det V_{AB}}\right)\right\}\,. \label{eq:sab}
\end{eqnarray}

Up to local unitary operations, any two-mode covariance matrix $V_{AB}$ can be written in a standard form where all the blocks $A,B,C$ of Eq.~(\ref{eq:cm}) are diagonal, with $A = \text{diag}(a,a), B=\text{diag}(b,b), C=\text{diag}(-c,d)$ \cite{Simon00,Duan00}. A particularly relevant class of Gaussian states is represented by two-mode squeezed thermal states, whose standard form is specified by three parameters only ($d=c$):
\begin{equation}\label{eq:sform}
V_{AB}=
\left(\begin{array}{cc|cc}
a & & -c & \\
& a & & c \\ \hline
-c & & b & \\
& c & & b
\end{array}\right)\,,
\end{equation}
subject to the bona fide condition (\ref{eq:bonafide}). This class includes the pure two-mode squeezed vacuum states, with covariance matrix $V_{AB}^{\rm TMS}(r)$ specified by \begin{equation}\label{eq:tms}
a=b=\cosh(2r)\,,\quad c=\sinh(2r)\,,
 \end{equation}
 where $r$ is a real squeezing parameter. These states approach the ideal EPR state \cite{EPR35} in the asymptotic limit of large squeezing $|r| \rightarrow \infty$.

\subsection{Gaussian channels}
Quantum channels which map Gaussian states into Gaussian states are known as Gaussian channels \cite{Caves1994,Braunstein2005,Wolf2007,Weedbrook2012,Serafini2017}. Up to additional displacements, their action on a $n$-mode continuous variable system can be described by two matrices $(X,Y)$, with $Y=Y^\top$, which transform first and second moments as follows,
\begin{equation} \label{GChannel}
d\mapsto Xd, \quad V \mapsto XVX^{\top}+Y,
\end{equation}
Gaussian channels must satisfy the complete positivity condition \cite{Serafini2017} $Y+iX\omega^{\oplus n} X^{\top}\geq i\omega^{\oplus n}$, which in the case of a single mode ($n=1$) reduces to:
$Y\ge0,\  \sqrt{\det Y}\geq|1-\det X|$.

\begin{figure*}[t!]
\center
\includegraphics[width=8cm]{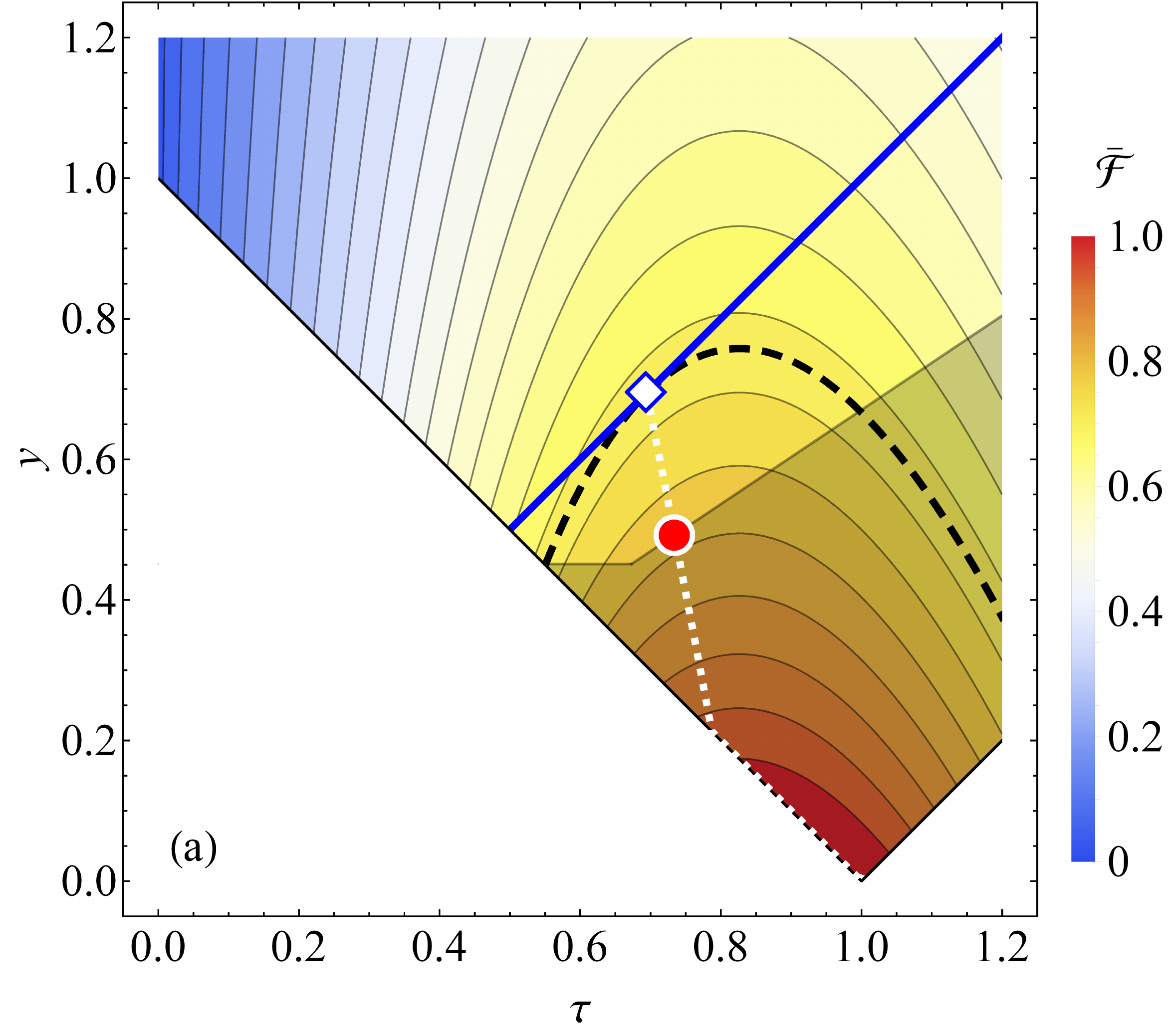}\hspace*{.5cm}
\includegraphics[width=8cm]{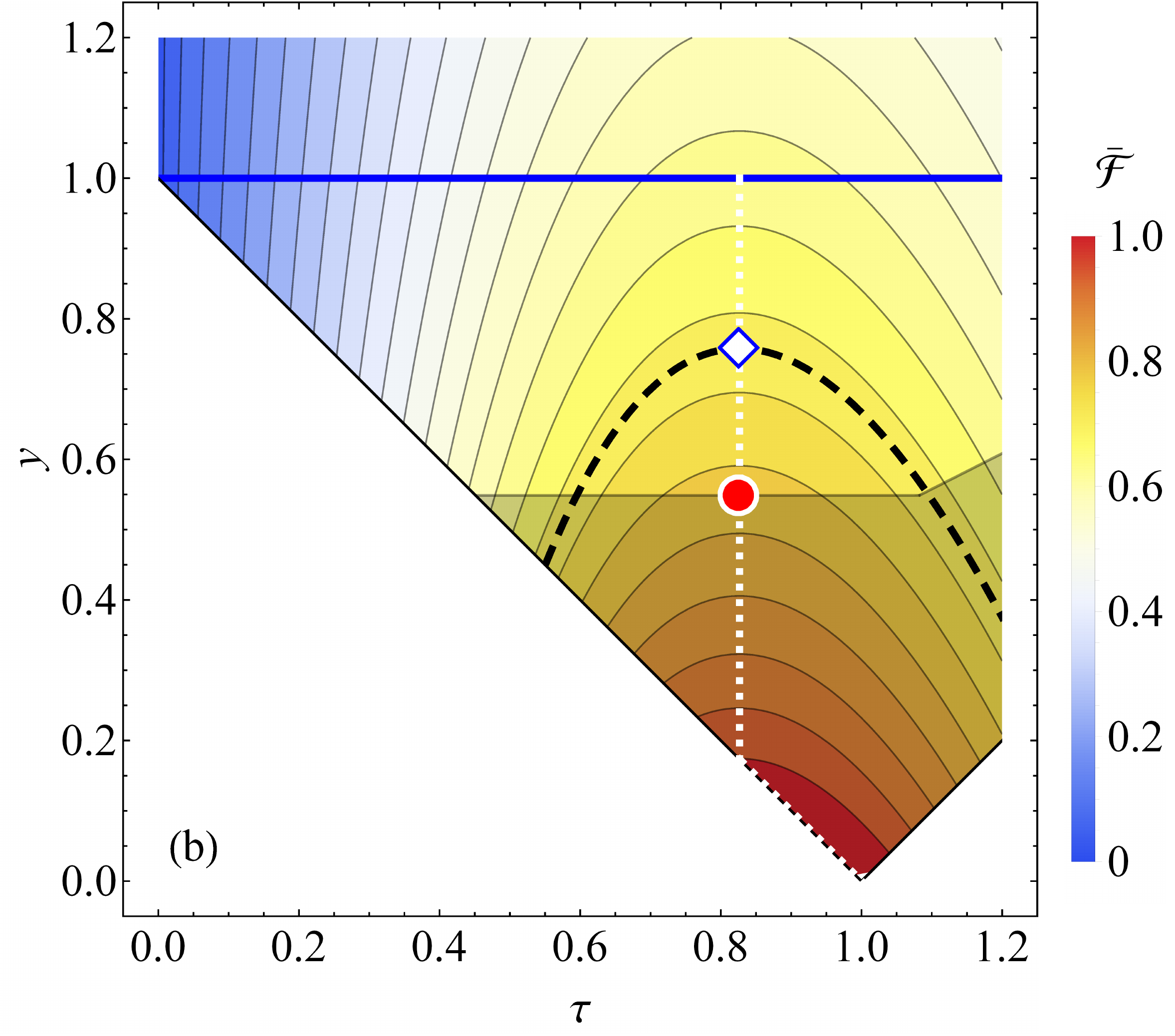}\\
    \caption{Diagram of single-mode phase-insensitive Gaussian channels defined by points in the $(\tau, y)$ plane according to Eq.~(\ref{pic}). The white area accommodates unphysical channels, which violate Eq.~(\ref{tauyCPT}). Points above the solid blue line in panel (a) correspond to $B \to A$ Gaussian steerability-breaking channels, defined by Eq.~(\ref{tauyBA}), while points above the solid blue line in panel (b) correspond to $A \to B$ Gaussian steerability-breaking channels, defined by Eq.~(\ref{tauyAB}). In both panels, the contour plot depicts the fidelity $\bar{\cal F}_{\lambda}(\tau,y)$, Eq.~(\ref{eq:fidelityxy}), between an input coherent state $\ket{\alpha}^{\rm in}$ and the corresponding output state obtained by the action of a phase-insensitive channel specified by $(\tau,y)$, averaged over the input distribution $p_\lambda(\alpha)$, with $\lambda=0.2$. The dashed black contour identifies the no-cloning threshold $\bar{\cal F}_{\lambda} = \bar{\cal F}_{\lambda}^{(s)}$, Eq.~(\ref{benchmark}); channels below such contour yield average fidelity exceeding the threshold, hence amounting to secure teleportation protocols. The dotted white line spans the family of optimal teleportation protocols at fixed (a) $B \to A$ Gaussian steerability $s_{ba}$ and (b) $A \to B$ Gaussian steerability $s_{ab}$, each starting from $0$ at the intersection with the solid blue line, increasing down the corresponding dotted white line, and converging towards the identity channel $(1,0)$ which is reached in the limit of infinite steerability. Relevant channels are highlighted as special points in the figure: The red-filled circle corresponds to the optimal channel with (a) $s_{ba}=0.4$ and (b) $s_{ab}=0.6$, while the white-filled diamond corresponds to the optimal channel at the boundary of the secure teleportation region, given by (a) $s_{ba}=0$ and (b) $s_{ab}=s_{ab}^{\min}$, Eq.~(\ref{minsab}).
    The shaded gray area corresponds to channels not accessible by teleportation schemes exploiting resources with finite steerabilities $s_{ba}$ and $s_{ab}$ according to Eq.~(\ref{bounds}), with (a) $s_{ba}=0.4$ and $s_{ab}$ defined by Eqs.~(\ref{stiff}) and (\ref{eq:tauba}), and (b) $s_{ab}=0.6$ and $s_{ba}$ defined by Eqs.~(\ref{stiff2}) and (\ref{eq:tauab}). All the quantities plotted are dimensionless.
\label{fig}}
\end{figure*}

We will devote our attention to single-mode phase-insensitive Gaussian channels, which model typical sources of noise in optical communications, and are described by
\begin{equation}
X=\sqrt{\tau} \mathbb{1}, \quad Y=y \mathbb{1},  \label{pic}
\end{equation}
where $\tau$ and $y$ are scalars playing the role of transmissivity (or gain) and added noise, respectively \cite{Schaefer2013}.
In the plane $(\tau,y)$, as illustrated in Fig.~\ref{fig}, single-mode phase-insensitive Gaussian channels can be then classified  into:
\begin{itemize}
\item completely positive if and only if \cite{Serafini2017}
\begin{equation}\label{tauyCPT}
y \ge |1- \tau|\,;
\end{equation}
\item entanglement-breaking if and only if \cite{Holevo2008}
\begin{equation}\label{tauyEB}
y \ge 1+ |\tau|\,;
\end{equation}
\item $B \to A$ Gaussian steerability-breaking if and only if \cite{Heinosaari2015}
\begin{equation}\label{tauyBA}
y \ge \frac12\left(1+ |2\tau-1|\right)\,;
\end{equation}
\item $A \to B$ Gaussian steerability-breaking if and only if \cite{Heinosaari2015}
\begin{equation}\label{tauyAB}
y \ge \max\left\{|1-\tau|,\,1\right\}\,.
\end{equation}
\end{itemize}
Here by an entanglement-breaking (resp.~steerability-breaking) channel we mean a single-mode channel defined by $(X,Y)$ such that, when applied to mode $B$ of any two-mode input state, produces always a separable (resp.~unsteerable) two-mode output state. In formula,
\begin{equation}\label{eq:breaking}
V_{AB}^{\rm out} = (\mathbb{1}_A \oplus X_B) V_{AB}^{\rm in} (\mathbb{1}_A \oplus X_B)^\top + (\mathbb{0}_A \oplus Y_B)
\end{equation}
fulfills  (\ref{eq:separable}), (\ref{eq:unsteerableba}), or (\ref{eq:unsteerableab}) for any $V_{AB}^{\rm in}$ when the channel is specified by (\ref{tauyEB}), (\ref{tauyBA}), or (\ref{tauyAB}), respectively. Notice that this holds in particular for the Choi state of the channel \cite{Giedke2002}, which is obtained when the input is the ideal EPR state, $V_{AB}^{\rm in} = V_{AB}^{\rm TMS}(r)$ [Eq.~(\ref{eq:tms})] in the limit $r \rightarrow \infty$.

Our analysis will restrict to phase-covariant channels, specified by $\tau  \geq 0$. The Gaussian channels on the lower boundary of the completely positive region, i.e.~with $(\tau,y)$ saturating (\ref{tauyCPT}), reproduce in particular the quantum limited amplifier for $\tau >1$ and the quantum limited attenuator for $0 \leq \tau <1$,  while $\tau=1$ identifies the identity channel, which corresponds to a perfect teleportation channel.

\subsection{Continuous variable teleportation}\label{sec:bk}

According to the BK teleportation protocol \cite{Braunstein1998}, Alice and Bob share a two-mode Gaussian resource state with vanishing displacement vector and covariance matrix $V_{AB}$ given by Eq.~(\ref{eq:cm}). Alice is supplied an unknown input single-mode Gaussian state with displacement vector $d^{\rm in}$ and covariance matrix $V^{\rm in}$; she mixes it with her mode $A$ of the resource state at a balanced beam splitter, and then performs a double homodyne detection. After Alice classically communicates the measurement results to Bob, he  performs a displacement on his mode $B$ with a tunable gain $g$ (see e.g.~\cite{Fiurasek2002,VanLoock2002} for details). At the end of the protocol, mode $B$ is transformed into the output state with  displacement vector and covariance matrix given by \cite{Liuzzo2017}
\begin{equation}\mbox{$d^{\rm out}= g\, d^{\rm in}$, \quad
 $V^{\rm out}=g^2V^{\rm in}+g^2  Z  A Z +g( Z  C+C^{\top}  Z )+B$}\,,\label{eq:bk}
 \end{equation}
 with $Z=\text{diag}(1,-1)$.
Comparing input and output one sees that, if the shared resource is a two-mode squeezed thermal state with covariance matrix $V_{AB}$ as in Eq.~(\ref{eq:sform}), then the BK teleportation protocol with gain $g$ amounts overall to the action of a single-mode phase-insensitive Gaussian channel, defined by \eqref{pic} with parameters
\begin{equation}
\tau=g^2\,,\quad  y= g^2 a -2 g c +b\,. \label{simul}
\end{equation}

\section{Teleportation simulation of Gaussian channels with finite steerability}\label{sec:sim}

In general, teleportation protocols can be used to simulate a variety of physical quantum channels. The problem of channel simulation by teleportation using entanglement as a resource, together with its wider implications for assessing the ultimate limitations of quantum communication, has received significant attention \cite{Giedke2002,Mari2008,Niset2009,Pirandola2015,Wilde2017,Liuzzo2017}.

 Here we are interested in a variation of this problem. Given an arbitrary pair of parameters $(\tau,y)$ describing a single-mode phase-insensitive Gaussian channel, constrained to (\ref{tauyCPT}), we want to find the two-mode Gaussian resource state described by a covariance matrix $V_{AB}$ which can be used in a continuous variable teleportation protocol to simulate the corresponding channel with a minimum amount of steerability, and possibly finite mean energy.

% We take in particular the $B \rightarrow A$ Gaussian steerability as our fixed resource of choice

For a fixed Gaussian steerability (in either direction) of the resource state shared by Alice and Bob, not all phase-insensitive channels can be implemented through the BK protocol described by Eq.~(\ref{eq:bk}). This follows from the fact that the degree of Gaussian steerability is monotonically nonincreasing under Gaussian local operations and classical communication \cite{Lami2016}, and cannot be distilled by means of such operations, similarly to what happens for entanglement \cite{Giedke2002}. Hence, let us suppose that Alice wants to transfer one mode of an ideal EPR state to Bob through a teleportation protocol which simulates the channel specified by the pair $(\tau,y)$. Let us also suppose that Alice and Bob share a two-mode Gaussian resource state with $V_{AB}$ given by Eq.~(\ref{eq:cm}) and with a fixed amount of Gaussian steerability from $B$ to $A$ [Eq.~(\ref{eq:sba})],  ${\cal S}_{B \to A}(V_{AB}) = s_{ba}$. We know that the Gaussian steerability of the output state ${\cal S}_{B \to A}(V_{AB}^{\rm out}) =\max\left\{0,\log(\tau/y)\right\}$ cannot exceed the steerability of the state initially shared by the two parties. Therefore, they can simulate the channel defined by $(\tau,y)$  only if $y\geq e^{-s_{ba}}\tau$. An analogous reasoning could be done for the Gaussian steerability from $A$ to $B$ [Eq.~(\ref{eq:sab})],  ${\cal S}_{A \to B}(V_{AB}) = s_{ab}$, obtaining that Alice and Bob can simulate the channel with $(\tau,y)$  only if $y\geq e^{-s_{ab}}$.

We have then that, given a resource state specified by a covariance matrix $V_{AB}$ with steerability degrees $s_{ba}$ and $s_{ab}$, the single-mode phase-insensitive Gaussian channels $(\tau,y)$ which could be implemented through a continuous variable teleportation protocol are those satisfying
\begin{equation}\label{bounds}
  (\tau,y)\quad\mbox{s.t.}\quad\begin{cases} y\geq e^{-s_{ba}}\tau \\ y\geq e^{-s_{ab}} \end{cases}.
\end{equation}
In particular,  any Gaussian channel may be implemented through a teleportation protocol which uses as a shared resource the Choi state of the channel itself \cite{Giedke2002,Mari2008,Niset2009,Pirandola2015,Wilde2017}. This state, however, has infinite mean energy and its implementation is thus impracticable in any experimental scenario.

We will show that it is possible to construct realistic classes of Gaussian resource states, with minimum steerability degrees (equal to the ones of the Choi state) and finite mean energy, able to simulate all phase-insensitive Gaussian channels saturating the boundaries in Eq.~(\ref{bounds}), with the only exclusion of the quantum limited attenuator and amplifier, for which these optimal states have again a diverging mean energy. These findings add to the recent analysis \cite{Liuzzo2017} where entanglement, rather than steering, was considered as a limited resource.
In the following, we investigate the two steering directions separately, and find in particular optimal states tailored to each of them.

\subsection{Optimal resources with fixed $B \to A$ steerability}\label{sec:cba}

We begin by discussing the case of a fixed $B \to A$ steerability, given by Eq.~(\ref{eq:sba}).
The optimal two-mode resource states for teleportation simulation with a finite $s_{ba}$ are found by starting from the covariance matrix $V_{AB}$ in the standard form given by Eq.~(\ref{eq:sform}), then fixing  $b$ such that ${\cal S}_{B \to  A}(V_{AB})=s_{ba}$, and $c$ such that the bound $y\geq e^{-s_{ba}}\tau$, with $y$ given by Eq.~\eqref{simul} together with $g=\sqrt{\tau}$, holds with equality:
\begin{align}
b =(a-e^{-s_{ba}})\tau\,, %\nonumber \\
 \qquad c =(a-e^{-s_{ba}})\sqrt{\tau}\,, \qquad
  a \geq\max\{a_+,a_-\}\,,\qquad\mbox{with}\quad a_{\pm}=\frac{e^{s_{ba}}+\tau(e^{-s_{ba}}\pm1)}{e^{s_{ba}}(\pm\tau \mp1)+\tau} \,.
   \label{coeff}
\end{align}
where the condition given on $a$ ensures that (\ref{eq:bonafide}) is fulfilled, i.e., that the state is physical. Within this class of states, we choose those with the minimal mean energy according to Eq.~(\ref{eq:energy}), given by the value of $a$ which saturates the inequality in \eqref{coeff}. The resulting states are two-mode asymmetric squeezed thermal states with a unit symplectic eigenvalue ($\nu^-_{AB}=1$), partially saturating the uncertainty principle (\ref{eq:bonafidenu}) \cite{Adesso2005a,Adesso2004}.
Notice that the lower bound on the coefficient $a$ (and hence on the mean energy since the latter scales linearly with $a$) diverges only at the points $(\tau=\frac{1}{1+e^{-s_{ba}}},\, y=\frac{1}{1+e^{s_{ba}}})$, corresponding to the quantum limited attenuator, and $(\tau=\frac{1}{1-e^{-s_{ba}}},\, y=\frac{1}{e^{s_{ba}}-1})$, corresponding to the quantum limited amplifier. For all intermediate values of $\tau$, that is, $\frac{1}{1+e^{-s_{ba}}}<\tau< \frac{1}{1-e^{-s_{ba}}}$, all channels along the boundary saturating the first inequality in \eqref{bounds} can be simulated using the states of Eq.~(\ref{coeff}) with a finite mean energy and minimum steerability ${\cal S}_{B \to  A}(V_{AB})=s_{ba}$.

On the other hand, notice that the steerability from $A$ to $B$ of the family of states of Eq.~\eqref{coeff} is given by
\begin{equation}\label{stiff}
  s_{ab}=-\log\left( \frac{e^{-2s_{ba}}(a\ e^{s_{ba}}-1)\tau}{a}\right)~,
\end{equation}
which is a decreasing function of the parameter $a$. This means that in the family of states of Eq.~\eqref{coeff}, for a fixed $B \to A$ steerability $s_{ba}$ the state with the least possible mean energy is the one with the maximum $A \to B$ steerability. Conversely, the state with the minimum steerability from $A$ to $B$, within the family of Eq.~(\ref{coeff}), has infinite mean energy. % and it is indeed the Choi state of the channel.  {\color{red}non sono completamente sicuro di questo claim!}

\subsection{Optimal resources with fixed $A \to B$ steerability}\label{sec:cab}

We now consider the case of a fixed $A \to B$ steerability, given by Eq.~(\ref{eq:sab}). We hence want to find a class of resource states, described by a covariance matrix $V_{AB}$ with ${\cal S}_{A \to B}(V_{AB})=s_{ab}$, which allow us to saturate the second inequality in \eqref{bounds}. Following a construction analogous to the one done in the previous subsection, we find
\begin{align}
 b =a\tau+e^{-s_{ab}}\,, %\nonumber \\
 \qquad c =a\sqrt{\tau} \,, \qquad
 a\geq{\rm max}\left\{a_+, a_-\right\}\,, \quad \mbox{with} \quad a_{\pm}=\left[{\tau\left(\frac{1}{e^{s_{ab}}\pm1}\mp1\right)\pm1}\right]^{-1}
  \,. \label{coeff2}
\end{align}
Also in this case, the state with the minimal mean energy is the one given by the value of $a$ which saturates the inequality in \eqref{coeff2}. Once more, the lower bound for the coefficient $a$ diverges only at the points $\left(\tau=1-e^{-s_{ab}},\,y=e^{-s_{ab}}\right)$  and $\left(\tau=1+e^{-s_{ab}},\, y=e^{-s_{ab}}\right)$, corresponding to the quantum limited attenuator and amplifier, respectively.

The steerability from $B$ to $A$ of the family of states defined by Eq.~\eqref{coeff2} reads
\begin{equation}\label{stiff2}
  s_{ba}=-\log\left(\frac{a}{a\ e^{s_{ab}} \tau +1}\right)~,
\end{equation}
which is a decreasing function of $a$. Analogously to the previous case, choosing the minimal $a$ is equivalent to maximise the steerability from $B$ to $A$, while, on the other hand, for $a\rightarrow\infty$ we have that $s_{ba}$ takes its minimum value and the simulated channel lies at the intersection of the boundary lines $y=e^{-s_{ab}}$ and $y=e^{-s_{ba}}\tau$, delimiting the region of implementable channels according to \eqref{bounds}.

\section{Optimal secure teleportation of coherent states with limited resources}\label{sec:opt}

Here we exploit the results of Sec.~\ref{sec:sim} in order to solve the concrete problem of determining the maximum average fidelity $\bar{\cal F}$ for secure teleportation of an alphabet of coherent states of light, given a finite steerability available as a resource.

Let us suppose that Alice wants to teleport to Bob a  coherent state \cite{Glauber1963} $\ket{\alpha}^{\rm in}$ with unknown amplitude $\alpha \in \mathbb{C}$ sampled from a Gaussian phase space distribution \begin{equation}\label{eq:plambda}
p_\lambda(\alpha)=\frac{\lambda}{\pi} e^{-\lambda|\alpha|^2}\,
\end{equation}
with finite variance $\lambda^{-1}$. According to Eq.~(\ref{eq:avf}), the average input-output teleportation fidelity is given by
\begin{equation}\label{eq:fcoh}
  \bar{\mathcal{F}}_\lambda=\int_\mathbb{C}d^2\alpha\ p_\lambda(\alpha)\ {}^{\rm in}\bra{\alpha}\rho^{\rm out}\ket{\alpha}^{\rm in}\,.
\end{equation}
As discussed in the Introduction, the security of the teleportation protocol is certified if the average fidelity $\bar{\mathcal{F}}_\lambda$ exceeds the so-called no-cloning threshold \cite{GG}, which for the considered input ensemble with inverse variance $\lambda \geq 0$ and in the assumption of Gaussian cloners is given by  \cite{Cochrane2004,GrosshansPhD}
\begin{equation}\label{benchmark}
  \bar{\mathcal{F}}^{(s)}_\lambda=
  \begin{cases}
  \displaystyle\frac{2(1+\lambda)}{3+\lambda}~,\quad&\lambda\leq\sqrt{2}-1\,;\\ \quad \\
  \displaystyle\frac{2\lambda}{3-2\sqrt{2}+2\lambda}~,\quad&\mbox{otherwise}\,.
  \end{cases}
\end{equation}
Achieving $\bar{\mathcal{F}}_\lambda > \bar{\mathcal{F}}_\lambda^{(s)}$ guarantees that no better copy of Alice's input state than Bob's output can be received by  Eve, who is entitled to use any Gaussian operation to intercept the communication between Alice and Bob.\footnote{In the limit of uniform input distribution ($\lambda \rightarrow 0$) the threshold in Eq.~(\ref{benchmark}) reduces to $\bar{\mathcal{F}}^{(s)}_0=2/3$ \cite{GG}, which is the best $1 \to 2$ single-clone fidelity achievable using Gaussian cloners. It is known that non-Gaussian cloners can lead to a slightly higher single-clone fidelity, given by $\approx 0.6826$ \cite{Navez2005}. However, to the best of our knowledge, the generalisation of this result to the case of a non-uniform input distribution of coherent states ($\lambda > 0$) has not been reported. In this work, we restrict our analysis to Gaussian operations, which can be efficiently implemented in quantum optics, both for the communicating parties Alice and Bob and for the potential eavesdropper Eve. We thus adopt Eq.~(\ref{benchmark}) as our reference benchmark for (Gaussian) secure teleportation.}

\begin{figure*}[t!]
\center
\includegraphics[width=8cm]{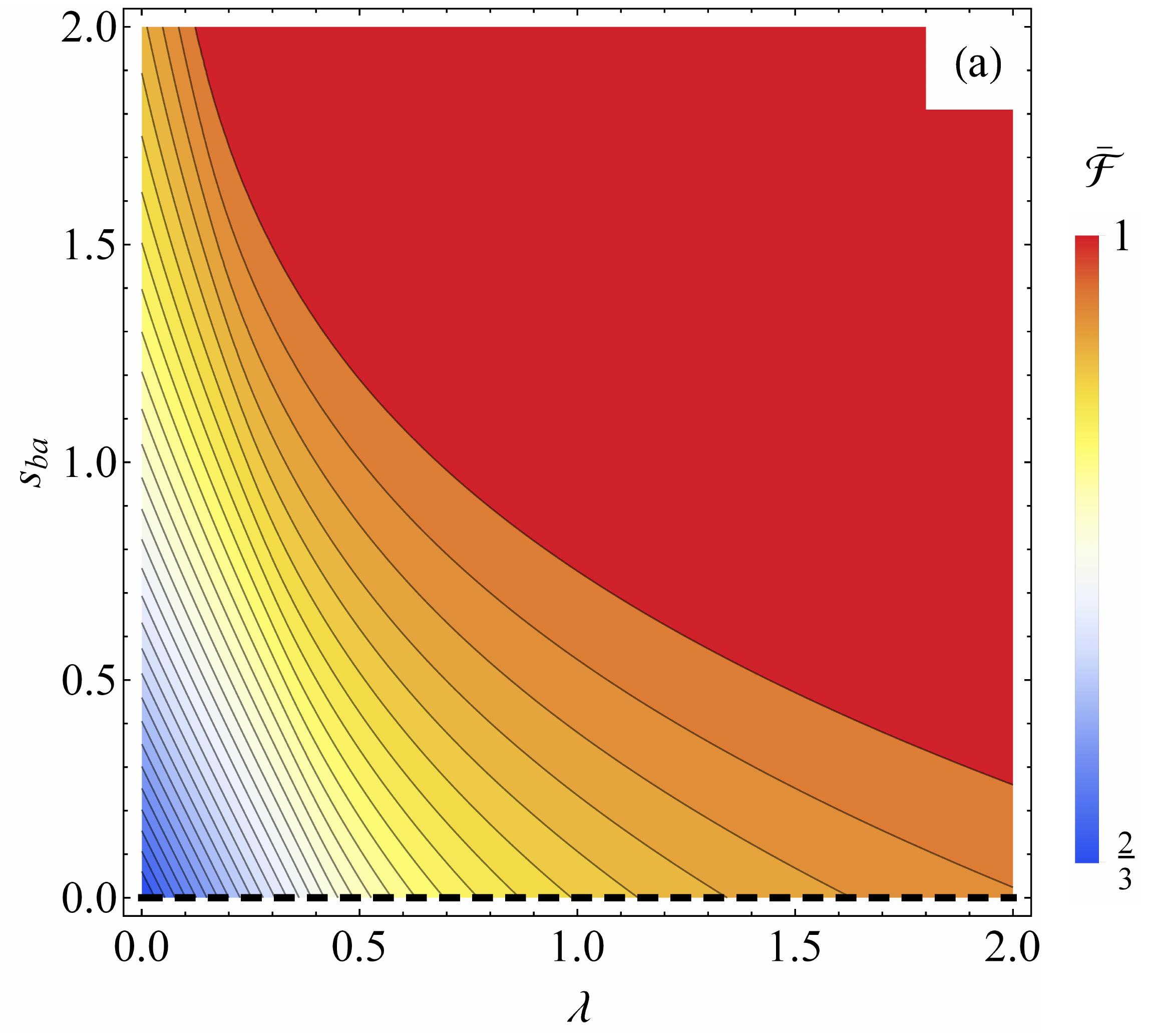}\hspace*{.5cm}
\includegraphics[width=8cm]{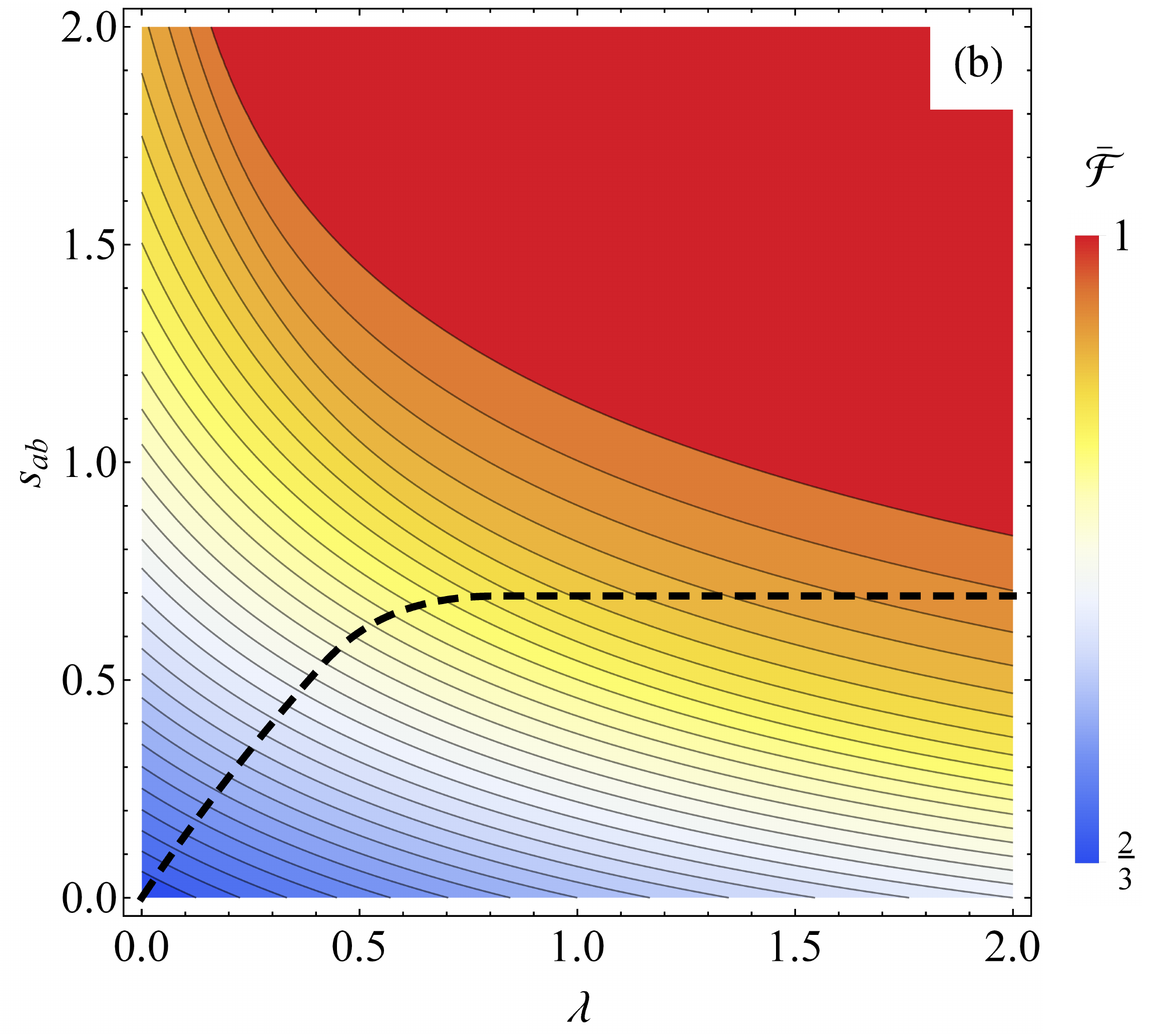}\\
    \caption{Contour plot of the optimal average fidelity $\bar{\cal F}^{\rm opt}_\lambda$ for teleporting an alphabet of coherent states of light sampled from a non-uniform phase space distribution $p_\lambda$ with variance $\lambda^{-1}$, when exploiting a resource with (a) fixed $B \to A$ Gaussian steerability $s_{ba}$, Eq.~(\ref{eq:fidelityba}), and (b) fixed $A \to B$ Gaussian steerability $s_{ab}$, Eq.~(\ref{eq:fidelityab}). The Gaussian teleportation is certified secure in the region above the dashed black line, in which case $\bar{\cal F}^{\rm opt}_\lambda$ exceeds the no-cloning threshold  ${\bar F}^{(s)}_\lambda$ given by Eq.~(\ref{benchmark}). This is achieved in (a) for any $s_{ba}>0$ , and in (b) for a finite $s_{ab}>s_{ab}^{\min}$, see Eq.~(\ref{minsab}). In the limit of uniform input distribution ($\lambda \rightarrow 0$) and vanishing steerability ($s_{ba}=s_{ab}=0$), the optimal average fidelity reduces to the benchmark ${\cal F}_0^{(s)} =  2/3$. All the quantities plotted are dimensionless
\label{fig2}}
\end{figure*}

We can study  the fidelity between the input state $|\alpha\rangle^{\rm in}$ and the output state obtained from the application of a phase-insensitive single-mode Gaussian channel with $(\tau, y)$,  averaged over the input distribution. Using Eq.~(\ref{eq:fcoh}), this yields \cite{Liuzzo2017}
\begin{equation}\label{eq:fidelityxy}
  \bar{\mathcal{F}}_\lambda(\tau,y)=\frac{2\lambda}{2(1-\sqrt{\tau})^2+\lambda(1+y+\tau)}~.
\end{equation}
The expression in Eq.~(\ref{eq:fidelityxy}) is overlayed in Fig.~\ref{fig} as a contour plot  in the plane $(\tau,y)$, together with the (dashed black) contour line defining the secure teleportation threshold, $\bar{\mathcal{F}}_\lambda(\tau,y)=\mathcal{F}_\lambda^{(s)}$ [Eq.~(\ref{benchmark})]. As shown in particular in Fig.~\ref{fig}(a), the latter line is tangent to the boundary $y=\tau$ of the $B \to A$ steerability-breaking region [Eq.~(\ref{tauyBA})] in the point  $\left((1+\lambda)^{-2},(1+\lambda)^{-2}\right)$, which corresponds to the best Gaussian teleportation channel that Alice and Bob can implement when sharing a resource with vanishing $s_{ba}$.

Thanks to the results of Sec.~\ref{sec:cba}, we can determine the optimal resource state, within the family of Eq.~(\ref{coeff}) at fixed  ${\cal S}_{B \to A}(V_{AB})=s_{ba}$, that Alice and Bob must share in order to maximise the average fidelity and beat the no-cloning threshold \eqref{benchmark}. As one can see in Fig.~\ref{fig}(a), the corresponding optimal teleportation channel is given for any $s_{ba}$ by the point at which the line $y=e^{-s_{ba}}\tau$, saturating the first boundary in (\ref{bounds}), is tangent to the corresponding contour of the average fidelity given by Eq.~(\ref{eq:fidelityxy}). Using simple geometry, the solution is hence given by
\begin{equation}\label{eq:tauba}
\tau^{\rm opt}_\lambda(s_{ba})=\max\left\{\frac{4e^{2s_{ba}}}{[\lambda+e^{s_{ba}}(2+\lambda)]^2},\, \frac{1}{1+e^{-s_{ba}}}\right\}\,,
\end{equation}
which yields the optimal average fidelity at fixed $s_{ba}$:
\begin{equation}\label{eq:fidelityba}
  \bar{\mathcal{F}}^{\rm opt}_\lambda(s_{ba})=
  \begin{cases}
  \displaystyle\frac{2 [\lambda+e^{s_{ba}}(2+\lambda)]}{2+\lambda+e^{s_{ba}}(4+\lambda)}~,
  \quad&\lambda\leq\frac{2 \left(\sqrt{e^{s_{ba}}(e^{s_{ba}}+1)}-e^{s_{ba}}\right)}{e^s_{ba}+1}
\,;\\ \quad \\
  \displaystyle\frac{\lambda  \left(e^{s_{ba}}+1\right)}{1+\lambda+e^{s_{ba}} \left(2+\lambda-2 \sqrt{e^{-s_{ba}}+1} \right)}~,
  \quad&\mbox{otherwise}\,.
  \end{cases}
\end{equation}

This confirms that the $B \to A$ steerability of the shared state, Eq.~(\ref{eq:sba}), is a meaningful, necessary and sufficient resource for optimal secure teleportation of coherent states of light: For any $\lambda$, the optimal average fidelity $\bar{\mathcal{F}}^{\rm opt}_\lambda(s_{ba})$ [Eq.~(\ref{eq:fidelityba})], plotted in Fig.~\ref{fig2}(a), is a monotonically increasing function of $s_{ba}$ and is larger than  $\mathcal{F}_\lambda^{(s)}$ [Eq.~(\ref{benchmark})] as soon as $s_{ba}>0$, reducing to the latter threshold exactly when $s_{ba}=0$. Notice further that the resource states used in the optimal protocol at fixed $s_{ba}$ also have a finite $A \to B$ steerability, given by Eq.~(\ref{stiff}). This is in agreement with the observation, originally made in the limiting case of uniform input distribution ($\lambda \rightarrow 0$), that secure teleportation of coherent states requires EPR steering in both directions \cite{He2015}.

For the sake of completeness, it is interesting to explore also the alternative scenario in which the $A \to B$ steerability ${\cal S}_{A \to B}(V_{AB})=s_{ab}$ in the resource state is fixed instead. From Fig.~\ref{fig}(b), we see that in this case the contour line
 $\bar{\mathcal{F}}_\lambda(\tau,y)=\mathcal{F}_\lambda^{(s)}$ does not intersect the boundary $y=1$ of the $A \to B$ steerability-breaking region [Eq.~(\ref{tauyAB})], which means that  $s_{ab}$ must have a finite value in order to overcome the no-cloning threshold \eqref{benchmark}. Indeed, using the results of Sec.~\ref{sec:cab}, it is easy to show that one must have
\begin{equation}\label{minsab}
  s_{ab}>s_{ab}^{\min} = \begin{cases}
    \displaystyle\log\Bigg(\frac{1}{2}(1+\lambda)(2+\lambda)\Bigg)~, \quad &0 \leq \lambda \leq \sqrt{2}-1\,; \\
    \displaystyle-\log \Bigg(\frac{\lambda}{\lambda+2}+\frac{3-2 \sqrt{2}}{\lambda}\Bigg)~, \quad &\sqrt{2}-1 < \lambda \leq 2(\sqrt{2}-1)\,; \\
    \displaystyle\log(2)~, \quad &\lambda>2(\sqrt{2}-1)\,,
    \end{cases}
\end{equation}
in order to achieve secure teleportation of coherent states of light. This means that, unless the input distribution is uniform \cite{He2015} ($\lambda \rightarrow 0$),  the $A \to B$ steerability of the shared state is a necessary but not sufficient resource for this task. Provided  \eqref{minsab} is satisfied, the optimal teleportation protocol Alice and Bob can implement uses a shared state belonging to the class of Eq.~\eqref{coeff2} with
\begin{equation}\label{eq:tauab}
\tau^{\rm opt}_\lambda(s_{ab})={\rm max}\left\{\frac{4}{(2+\lambda)^2},\,1-e^{-s_{ab}}\right\}\,.
\end{equation}
This leads to the corresponding optimal average fidelity at fixed $s_{ab}$:
\begin{equation}\label{eq:fidelityab}
  \bar{\mathcal{F}}^{\rm opt}_\lambda(s_{ab})=
  \begin{cases}
  \displaystyle\frac{2 e^{s_{ab}}(2+\lambda)}{2+\lambda+e^{s_{ab}}(4+\lambda)}~,
  \quad&\lambda\leq 2 \left(\sqrt{\frac{e^{s_{ab}}}{e^{s_{ab}}-1}}-1\right)
\,;\\ \quad \\
  \displaystyle\frac{\lambda}{\lambda + \left(\sqrt{1-e^{-s_{ab}}}-1\right)^2}~,
  \quad&\mbox{otherwise}\,.
  \end{cases}
\end{equation}
For any $\lambda$, the optimal average fidelity $\bar{\mathcal{F}}^{\rm opt}_\lambda(s_{ab})$ [Eq.~(\ref{eq:fidelityab})], plotted in Fig.~\ref{fig2}(b), is a monotonically increasing function of $s_{ab}$ and is larger than  $\mathcal{F}_\lambda^{(s)}$ [Eq.~(\ref{benchmark})] as soon as $s_{ab}>s_{ab}^{\min}$, reducing to the latter threshold exactly when $s_{ab}=s_{ab}^{\min}$ [Eq.~(\ref{minsab})]. Concerning steerability from $B$ to $A$, notice that the resource states of Eq.~(\ref{coeff2}) always have $s_{ba}>0$ when (\ref{minsab}) holds. More generally, ${\cal S}_{A \to B}(V_{AB})>\log(2)$ implies ${\cal S}_{B \to A}(V_{AB})>0$ for any two-mode Gaussian state with covariance matrix $V_{AB}$ \cite{Kogias2014}, confirming once again that two-way steerability is required for a certified secure teleportation of coherent states of light.

\section{Conclusions}
We determined the optimal Gaussian protocols for {\it secure} quantum teleportation\cite{GG} of an input alphabet of coherent states of light with generally non-uniform phase space distribution, exploiting Gaussian EPR steering\cite{Wiseman2007,Kogias2014} as a limited resource. Our analysis complements a recent study which focused instead on entanglement as a limited resource \cite{Liuzzo2017}, and goes beyond another recent investigation on the role of EPR steering in secure teleportation, which only specialised to uniform distributions of input coherent states \cite{He2015}. To obtain our results, we characterised the class of single-mode phase-insensitive Gaussian channels which can be simulated by teleportation exploiting a finite degree of steerability from $B$ to $A$ or from $A$ to $B$, identifying families of two-mode Gaussian resource states which are optimal for such a task and can be realised with finite mean energy (except when simulating a quantum limited attenuator or amplifier).

Our study confirms that steerability in both directions is required  to beat the benchmark for secure teleportation.\cite{GG,Cochrane2004,GrosshansPhD} From a practical pespective, we remark that in both cases analysed in this paper, i.e.~with fixed $B \to A$ or $A \to B$ Gaussian steerability as a limited resource, the optimal average teleportation fidelity $\bar{{\cal F}}^{\rm opt}$ defined respectively by Eq.~(\ref{eq:fidelityba}) or Eq.~(\ref{eq:fidelityab}) can be attained by a conventional BK teleportation protocol [Eq.~(\ref{simul})] with optimal gain $g^{\rm opt} = \sqrt{\tau^{\rm opt}}$, where $\tau^{\rm opt}$ is defined by Eq.~(\ref{eq:tauba}) or Eq.~(\ref{eq:tauab}), respectively.

In the near future, it would be desirable to implement the optimal continuous variable teleportation schemes devised here and hence to demonstrate experimentally the key role of Gaussian EPR steering for secure teleportation of coherent states of light with non-uniform phase space distribution. This would constitute a milestone for the quantum internet \cite{qinternet}, together with the recent optical demonstration of one-sided device independent quantum key distribution also exploiting Gaussian EPR steering \cite{Walk}.

\acknowledgments     %>>>> equivalent to \section*{ACKNOWLEDGMENTS}

We acknowledge discussions with Vittorio Giovannetti, Fr\'ed\'eric Grosshans, Qiongyi He, Ludovico Lami, Riccardo Laurenza, Andrea Mari, Ladislav Mi\v{s}ta Jr., Stefano Pirandola, and Margaret Reid.
This work was supported by the European Research Council (ERC) under the Starting Grant GQCOP (Grant No.~637352).

%%%%%%%%%%%%%%%%%%%%%%%%%%%%%%%%%%%%%%%%%%%%%%%%%%%%%%%%%%%%%
%%%%% References %%%%%

\bibliographystyle{apsrevfixedwithtitles}   %>>>> makes bibtex use spiebib.bst
\bibliography{biblioteleport}   %>>>> bibliography data in report.bib

\end{document}